\documentclass[aps,prd,onecolumn,groupedaddress]{revtex4}

\usepackage{graphicx}

\begin{document}


\title{Derivation of the TCP Theorem
       using Action Principles} 

%

\author{Mark Selover}
\email[]{selover@physics.utexas.edu}
\author{E.\ C.\ G.\ Sudarshan}
\affiliation{Center for Complex Quantum Systems, The University of Texas at Austin, Austin, Texas 78712, USA}
%


\date{22 Aug, 2013}

\begin{abstract}
We present a new derivation of the proof for the TCP/CPT Theorem using
the Dynamical Principle and Variation of Action methods first defined by Schwinger
in 1951 \cite{Schwinger-QFT1}. This new proof will significantly extend the TCP Theorem beyond
the original proofs by Pauli-L{\"{u}}ders and Jost,
which were significantly constrained by limits of free quantum fields
and the asymptotic condition. 
This paper gives a more fundamental proof based on invariance of the Variation of Action
with interactions included and no free field or asymptotic condition on the quantum fields.
This new proof for TCP/CPT can now be applied to more complicated 
quantum field systems that include n-particle bound states and unstable states.
\end{abstract}


\maketitle


\section{Introduction}
We present a new derivation of a proof of
the TCP/CPT Theorem using Schwinger's extended Action principle and Green's function methods.
Sixty years ago, Schwinger reformulated Quantum Field theory in terms of a 
generalized Variation of Action Principle (first defined in detail in \cite{Schwinger-QFT1}),
and showed how propagators and multiparticle Green's functions
can be derived from this Action principle in
\cite{Schwinger-Green1}, \cite{Schwinger-Green2}, and \cite{Schwinger-Gauge1}.
Schwinger first described the concept of ``Strong Reflection''
for TCP in relativistic quantum fields in \cite{Schwinger-QFT1}.
He also described the critical connection between
T inversion and C (complex/charge) conjugation in \cite{Schwinger-QFT2}.
T inversion reverses the sign in front of the Action integral
and C conjugation is required to counterbalance.
The use of the anti-automorphism (operators with $-i$) on the interchange of the initial
and final spacelike surfaces therefore involves the conjugate algebra and C.
This critical connection between T and C
was described by Schwinger in a very short set of comments
(on page 720 in \cite{Schwinger-QFT2}).
This was done for a limited set of cases and as an intermediate  
step only in making a new proof for the Spin Statistics theorem
(Schwinger reviewed this connection between Spin Statistics and TCP again
in \cite{Schwinger-PNAS-1958}). 
Such a need for time reversal and the endpoint variations for field theories
was also postulated by P. Weiss in \cite{Weiss1}.
This critical connection between T and C 
emphasizes the importance of T, and so the original TCP name
will be used in this paper instead of the current CPT convention.

The derivation of the TCP theorem here will extend and generalize
Schwinger's definition of ``Strong Reflection'' and
comments on the key connection between T inversion and
C (complex/charge) conjugation
to a more general proof of TCP that can be applied to
interacting quantum fields involving bound states and unstable states. 

This general derivation of TCP
is consistent with the standard formulations of CPT,
which require the asymptotic condition for free quantum field cases
(see derivation in Weinberg \cite{Weinberg1} which uses an
an updated version of the Pauli-L{\"{u}}ders TCP proof \cite{Luders1} \cite{Luders2} \cite{Pauli1}).
The established Pauli-L{\"{u}}ders and Jost proofs for TCP
are based on S-matrix methods and
the constraints of asymptotic limits applied to free relativistic quantum fields
with no interactions,
which then allows the symmetries of the
complex Lorentz group to be used
to prove TCP Symmetry.
These assumptions break down and are not valid if the quantum fields are non-relativistic
or involve bound state or unstable state systems.

These significant limitations of the Pauli-L{\"{u}}ders and Jost proofs
was first seriously challenged by Kobayashi and Sanda in 1992 \cite{Kobayashi1},
who pointed out that these proofs fail to handle QCD quark bound states,
and had thus become obsolete and out of date.
Our extended proof of TCP here solves this problem by using the Variation of Action principle
which is applicable to a significantly larger domain of cases in quantum field theory.

\section{Review of History of Action Principle Methods}
\subsection{Action Principle in Classical Dynamics and Quantum Mechanics}
The Action Principle came into Analytical Mechanics more then a century ago
as the Principle of Least Action, which says that the actual trajectory of a particle
subject to arbitrary forces is one that makes the ``Action'' an extremum.
In classical dynamics, the Lagrangian density and space-time integral
the Action Functional have been used for a very longtime.
The principle of least Action, where the variation of the action between
fixed initial and final times give the equations of motion and the trajectory.
The initial and final values of the dynamical variables yield complete variation
as defined by Hadamard.
As long as the dynamical trajectory is a ``path'', the action integral is 
a ``path integral''.
 
The equations of motion and the Poisson bracket relations of classical mechanics
could be derived from a classical Action principle,
but to derive the Poisson bracket relations and the expression for the
Hamiltonian as the generator of true translations could only be
derived from a generalized Action principle formulation of mechanics.
In the generalized Action principle one considers a ``complete variation''
involving the change in the boundaries of integration.

When the Variation of the Action is enlarged to contain End point variations
of the dynamical variables and the time instants, $\delta t$ also changes
(from Hadamard).
The dependence on the limits leads to the notion of pairs of conjugate
variables, including the total Hamiltonian and the momenta (in Weiss \cite{Weiss1}).
When the dynamical system is a space-time field, one gets the momenta conjugate
to the space time coordinates.

But this was in the context of a classical theory.
When applied to quantum theory, instead of
considering the values of the dynamical variables,
following Schwinger, one can consider the custom matrix element
between the (possibly variable) limits and a change
in the transition matrix element 
between the initial and final limits
is related to the matrix element of the Action.

\subsection{Action Principle in Quantum Fields} 
When these connections are extended to the 
quantum theory of fields (W. Heisenberg and W. Pauli),
we can develop a quantum theory of fields.
The bounding variations satisfy quantum Poisson brackets in accordance
with the formula of Dirac (in \cite{Dirac1})
$\{ w_{1}, w_{2} \} = \frac{i}{\hbar}(w_{1} w_{2} - w{2} w{1})$
and emphasized by P. Weiss (in \cite{Weiss1}).

If the dynamical variables possess more labels then just here, the Action integral
is a multidimensional ``field theory'', either classical or quantum.
The basic quanttity of interest become field equations.
The most important such systems are quantum fields which depend
parametrically on space-time points.
The corresponding equations become field equations in spacetime.

In a quantum theory the quantum action is an operator and one is led to evaluate
the matrix elements of the extended action between the initial and final states
of the system which for small variations would be a phase.
From these follow operator equations of
motion, commutation relations and dynamical equations of four-momenta 
and six-angular momenta.

In the traditional classical dynamical equations the Lagrangian density is a
function of spatial coordinates and or its time derivative,
but in the more advanced treatments the action density is a soluble operator
and the action itself is on operator functional and its matrix elements
between the intial and final spacelike surfaces is a pure imaginary phase.

In the hands of Schwinger, who invented the Green's function theory of quantized fields,
the basic starting point is the variational statement (which will be explained
in more detail in next section):
\begin{displaymath}
  \delta \left< \sigma_{i} | \sigma_{f} \right>
        = i \left< \sigma_{i} | \delta W | \sigma_{f} \right>
\end{displaymath}

To the extent that the theory deals with quantum fields, the action functional is explicitly Poincarre
invariant, and usually given in terms of Poincarre invariant products of the finite component
tensor or spinor amplitudes which retain the Poincarre invariance in an explicit manner.

\subsection{Review of Schwinger's Dynamical Principle and Variation of Action Methods}
Schwinger took up the general problem of the Action formulation of
Quantum Mechanics and Quantum Field Theory begining in the period 1947 to 1953.
We summarize here below Schwinger's methods of the Dynamical Priciple and Variation of Action,
defined in detail by Schwinger in \cite{Schwinger-QFT1} and \cite{Schwinger-QFT2},
and the 1-particle and 2-particle Green's functions derived by Schwinger in
\cite{Schwinger-Green1}, \cite{Schwinger-Green2}, \cite{Schwinger-Gauge1}.
A very good review of these articles is given by Schweber in \cite{Schweber1}.

In Schwinger's original definition (in \cite{Schwinger-QFT1} and \cite{Schwinger-QFT2}),
with the primary emphasis on 
simple 1 and 2 particle cases,
the quantum Action $W$ (in Schwinger's notation) was defined as follows:
\begin{equation}
   W_{1 2} = \int_{\sigma_{1}}^{\sigma_{2}} d^{4}x \; {\cal L}(x)
\end{equation}
where $\sigma_{1}$ and $\sigma_{2}$ are two space-like boundary surfaces
at the initial (subscript 1) and final (subscript 2) surface of a space-time volume $\Omega$,
$x$ is a localized space-time point on the surface or in the volume $\Omega$,
and ${\cal L}(x)$ is the Lagrangian density at $x$. 
The infinitesimal variation of the field $\psi$ at or inside the boundary surfaces
$\sigma_{1}$ and $\sigma_{2}$ gives the following fundamental Dynamical principle.
\begin{equation}
  \delta \left< \psi_{1},\sigma_{1} | \psi_{2},\sigma_{2} \right> = 
    i \left< \psi_{1},\sigma_{1} | \delta W_{1 2} | \psi_{2},\sigma_{2} \right>
\end{equation}
in which the endpoint variations $\delta\psi_{1}$, $\delta\psi_{2}$ are included.
The variation of the Action $\delta W_{1 2}$ is then defined as
\begin{equation}
   \delta W_{1 2} = \delta \int_{\sigma_{1}}^{\sigma_{2}} d^{4}x \; {\cal L}(x) 
\end{equation}
where the full variation of the Action depends on the endpoint variations $\delta\psi_{1}$, $\delta\psi_{2}$ 
on the boundary surfaces $\sigma_{1}$ and $\sigma_{2}$, and variation
of the lagrangian $\delta {\cal L}(x)$ in the interior of the volume bounded
by these surfaces.
This full variation of the Action can be defined here as:
\begin{displaymath}
         \delta W_{1 2} = \int_{\sigma_{2}}^{\sigma_{1}} d^{4}x \; \partial_{\mu} G_{\mu}(x) + 
                          \int_{\sigma_{2}}^{\sigma_{1}} d^{4}x \; \delta \left( {\cal L}(x) \right)
                        = G_{2} - G_{1} + \int_{\sigma_{2}}^{\sigma_{1}} d^{4}x \; \delta \left( {\cal L}(x) \right)
\end{displaymath}
where $G_{2}$ and $G_{1}$ represent the numerical value for variation at the boundary surfaces.
\subsection{Generalize the Variation of Action principle to n-particle systems}
In the quantization of a system of $n$ particles to a general quantum field,
the state of the system is defined then as follows:
\begin{displaymath}
   \Psi  =
        \frac{1}{\sqrt{n!}} \int d^{4}x_{1} \dots d^{4}x_{n} \; \Psi(x_{1}, \dots , x_{n}) \;
                  \psi^{\dagger}(x_{1}) \dots \psi^{\dagger}(x_{n}) | 0 >
\end{displaymath}
The quantum Action, using the standard convention $S$ here, is then defined as:
\begin{equation}
   S_{if} = \int_{\sigma_{i}}^{\sigma_{f}} d^{4}x \; {\cal L}(\Psi, \partial_{\mu}\Psi)
\end{equation}
where the subscripts $i$ and $f$ represent initial and final states. 
This is true for the quantum field of any general $n$-particle system.

The infinitesimal variation of the field $\Psi$ at the boundary surfaces
$\sigma_{i}$ and $\sigma_{f}$ is given again by the endpoint variations 
$\delta\Psi_{i}$ and $\delta\Psi_{f}$.
With the quantum Action defined in this way for a general n-particle system,
the fundamental Dynamical principle then becomes:
\begin{equation}
  \delta \left< \Psi_{i},\sigma_{i} | \Psi_{f},\sigma_{f} \right> = 
    i \left< \Psi_{i},\sigma_{i} | \delta S_{i f} | \Psi_{f},\sigma_{f} \right>
\end{equation}
in which the endpoint variations $\delta\Psi_{i}$, $\delta\Psi_{f}$ are included.
The variation of the Action is then defined as:
\begin{equation}
   \delta S_{i f} = \delta \int_{\sigma_{i}}^{\sigma_{f}} d^{4}x \;
                       \left( {\cal L} (\Psi, \partial_{\mu}\Psi) \right)
\end{equation}
\subsection{Manifest Covariance and the Connectivity of the Full Poincare Group}
The concept of ``Strong Reflection'' TCP for the combined operations of
time inversion T, charge conjugation C, and parity P
was initially described by Pauli as a key intermediate step of his Non-relativistic theory for
Spin-Statistics Connection. This concept of ``Strong Reflection'' was then more fully
developed by Schwinger for Relativistic Quantum field theory using the
Action principle described in \cite{Schwinger-QFT1}, and then extended 
in his development of a proof for Spin-Statistics Connection in \cite{Schwinger-QFT2}.
For the operations of P parity
(where $x_{j} \rightarrow x^{'}_{j} = -x_{j} \; (j=1,2,3)$),
and T time inversion (where $x_{0} \rightarrow x^{'}_{0} = -x_{0}$)
in the combined operation of ``Strong Reflection'',
the elements of the proper Lorentz real sub-group $SO(3,1)$
cannot reach this continuously from the identity,
but there are real elements which make this possible
in the complex Lorentz group $L({\cal C})$. 
\subsection{Entangling T time inversion and C complex conjugation }
The effect of T time inversion has the consequence of
interchanging the positions of initial and final states in the Action integral
which changes the sign in front of the action integral.
The change of sign of the Action 
then changes the sign of $i$ in the Dynamical principle
as follows:
\begin{eqnarray}
  T\left( \delta \left< \Psi_{f},\sigma_{f} | \Psi_{i},\sigma_{i} \right> \right)
  & = & T \left( i \left< \Psi_{f},\sigma_{f} | \delta S_{f i} | \Psi_{i},\sigma_{i} \right> \right)  \nonumber \\
  & = & T \left( i \left< \Psi_{f},\sigma_{f} \left| 
             \delta \int_{\sigma_{f}}^{\sigma_{i}} d^{4}x \; {\cal L} (\Psi, \partial_{\mu}\Psi) 
          \right| \Psi_{i},\sigma_{i} \right> \right)  \nonumber \\
  & = & - i \left< \Psi_{i},\sigma_{i} \left|
                      \delta \; \int_{\sigma_{i}}^{\sigma_{f}} d^{4}x \; {\cal L}(\Psi, \partial_{\mu}\Psi) \;
                   \right| \Psi_{f},\sigma_{f} \right>
\end{eqnarray}
Some other explicit change in the Action integral must be done to compensate for this.

For the representations of Relativistics Quantum fields such as
the Klein-Gordon, Dirac, Weyl, and Maxwell field, where both positive and negative energy
states are implicit in the 
covariant fields, T time inversion reverses particle paths
when the boundary surfaces $\sigma_{i}$ and $\sigma_{f}$ are interchanged,
forcing an implicit inversion to negative energy states.
This demands that the dynamical variables undergo an anti-automorphism C
which involves the strong time reversal TC instead of the weak (Wigner) time reflection T
for all fields.

The operation of T time inversion in the Dynamical principle in eqn (8) 
manifests as a change in sign of the Action integral, which propagates
to change the sign of i in front. 
This change in sign must now be compensated for by another
change in sign in the Action integral to maintain overall
TCP Invariance after Strong reflection.
To compensate for this, Schwinger used (in \cite{Schwinger-QFT2} on pg.\ 720),
the Complex Conjugation operation to the algebra on dynamical variables in $\cal L$.
This is the same operation as in Charge Conjugation C for Scalar and Vector fields. For Dirac fields
in the standard representation, the C operation introduces phase factors, but for this part
of the discussion we will choose the Majorana representation 
\footnote{In the Majorana representation of a Dirac field, the C operation produces no phase factors}
so that the C operation only involves the Complex Conjugation operation.
The criterion then to maintain Invariance of the full TCP Strong Reflection operation
after T inversion changes the sign of the Action integral in eqns (3) and (11), and overall changes the sign  
in front of $i$ in the Dynamical principle in eqns (2) and (10), is to perform the
C Charge Conjugation operation on the basic fields in the Lagrangian.
This simplifies to Complex Conjugation for all basic fields in the Lagrangian as follows 
\begin{eqnarray}
   C \left( {\cal L}( \Psi, \partial_{\mu}\Psi) \right) & \rightarrow &
          {\cal L}( i \Psi, i\partial_{\mu} \Psi)^{*}  \nonumber \\
          & \; = & - {\cal L}( i \Psi^{*}, i \partial_{\mu}\Psi^{*})
\end{eqnarray}
The operations of T time inversion and C charge conjugation
are then coupled in the operation TC to maintain proper invariance 
of the phase of the Action, and this becomes the strong condition on
T (that is described by Schwinger in \cite{Schwinger-QFT2} on pg.\ 720).

Other authors have noted this strong connection between T and C.
In a review of T time reversal in field theory by J.\ S.\ Bell \cite{Bell1}
a description of this difference between the Wigner weak condition on T inversion
(T only), and the Schwinger strong condition on T inversion (T and C are tighly coupled)
is given.
An even more explicit statement on the connection between T and C is given by
Dewitt (in \cite{Dewitt2} on page 729) that C is automatically part of T inversion.
Dewitt argues that the CPT theorem is essentially the PT theorem with C coming
as an integral part of T inversion for even dimension spacetime.

\subsection{Complete Derivation of TCP Invariance in Variation of the Action}
The fundamental statement and derivation then of TCP Invariance using the 
Dynamical Principle and Variation of Action method,
in the most general terms for a wide range of Quantum Field theories,
and not just limited to relativistic Lorentz invariant Quantum Fields where
TCP symmetry from the Complex Lorentz group is inherent. 
Then using the key definitions in eqns (4), (5), (6), (7) and (8),
the derivation is as follows:
\begin{widetext}
 \begin{eqnarray} 
  TCP \left( \delta \left< \Psi_{f},\sigma_{f} | \Psi_{i},\sigma_{i} \right> \right) 
    & = & TCP \left( i \left< \Psi_{f},\sigma_{f} | \delta S_{f i} | \Psi_{i},\sigma_{i} \right> \right) \nonumber \\
    & = & TCP \left( i \left< \Psi_{f},\sigma_{f} \left| \;
                                     \delta \left( \int_{\sigma_{f}}^{\sigma_{i}} (d^{4}x) 
                                            \left( {\cal L}(i \Psi, i\partial_{\mu}\Psi) \right)  \;
                                     \right) \right| 
                              \Psi_{i},\sigma_{i} \right> \right) \nonumber \\
    & = & i \left( \left< \Psi_{i},\sigma_{i} \left| \; 
          TCP \left( \delta \left( \int_{\sigma_{f}}^{\sigma_{i}} (d^{4}x)
                     \left( {\cal L}(i \Psi, i\partial_{\mu}\Psi) \right) \; 
              \right) \right) \right| \Psi_{f},\sigma_{f} \right> \right) \nonumber \\
    & = & i \left( \left< \Psi_{i},\sigma_{i} \left| \; 
                     \delta \left( - \int_{\sigma_{i}}^{\sigma_{f}} (d^{4}x) \;
                     TCP \left( \left( {\cal L}(i \Psi, i\partial_{\mu}\Psi)\right) \right) \; \right)
                   \right| \Psi_{f},\sigma_{f} \right> \right) \nonumber \\
    & = & i \left(  \left< \Psi_{i},\sigma_{i} \left| \;
          \delta \left( - \int_{\sigma_{i}}^{\sigma_{f}} (d^{4}x) \left( {\cal L}(i \Psi, i\partial_{\mu}\Psi)\right)^{*} \;
          \right) \right| \Psi_{f},\sigma_{f} \right> \right) \nonumber \\
    & = & -i \left< \Psi_{i},\sigma_{i} \left| \;
             \delta \int_{\sigma_{i}}^{\sigma_{f}} (d^{4}x) \; \left( {\cal L}(i\Psi, i\partial_{\mu}\Psi)\right)^{*} \;
             \right| \Psi_{f},\sigma_{f} \right> \nonumber \\
    & = & -i \left< \Psi_{i},\sigma_{i} \left| \;
             \delta \int_{\sigma_{i}}^{\sigma_{f}} (d^{4}x) \;
                  \left( - {\cal L}((i\Psi)^{*}, i\partial_{\mu}(\Psi)^{*} )\right) \;
             \right| \Psi_{f},\sigma_{f} \right> \nonumber \\
    & = & -i \left< \Psi_{i},\sigma_{i} \left| \;
           - \delta \int_{\sigma_{i}}^{\sigma_{f}} (d^{4}x) \; 
                  \left( {\cal L}( i\Psi^{*}, i\partial_{\mu}\Psi^{*} )\right) \;
             \right| \Psi_{f},\sigma_{f} \right> \nonumber \\
    & = & i \left< \Psi_{i},\sigma_{i} \left| \;
            \delta \int_{\sigma_{i}}^{\sigma_{f}} (d^{4}x) \left( {\cal L}( i\Psi^{*}, i\partial_{\mu}\Psi^{*}) \right) \;
            \right| \Psi_{f},\sigma_{f} \right> \nonumber \\
    & = & \delta \left< \Psi_{i},\sigma_{i} | \Psi_{f},\sigma_{f} \right>
 \end{eqnarray}
\end{widetext}
The test for TCP invariance of the variation of the Action is then
focused on TCP invariance of the  
Lagrangian density ${\cal L}( \Psi^{*}, \partial_{\mu}\Psi^{*} )$
where the n-particle field $\Psi$ has already gone through complex conjugation,
and TCP invariance of the variaton of the endpoints $\sigma_{i}$ and $\sigma_{f}$.
If the Lagrangian ${\cal L}( \Psi^{*}, \partial_{\mu}\Psi^{*} )$ and the
endpoint variations $\sigma_{i}\Psi_{i}$ and $\sigma_{f}\Psi_{f}$ are invariant under TCP, then the
Variation of the Action as a whole will be invariant under TCP as well.
All other parts of the Dynamical principle are now invariant under TCP operation.

This proof is then true for a large range of Quantum Field theories and generalized Lagrangian ${\cal L}$,
including Non-relativistic Quantum Fields which are disconnected from the inherent symmetries 
of the Complex Lorentz group.
For most Relativistic Quantum Fields which are Manifestly Lorentz Invariant
the Variation of the Action will be automatically invariant under TCP operations here
by definition, and the Action principles described here do not change this.
The purpose here is to widen the scope to include other Quantum Field theories where
the automatic connection to TCP symmetry in the Complex Lorentz group is not implicit, 

\subsection{TCP Invariance as a wider domain Action Principle}
The purpose here in this derivation of TCP as an Action Principle
is to widen the scope to include other Quantum Field theories where
the automatic connection to TCP symmetry in the Complex Lorentz group is not implicit,
As described in the last section, Relativistic Quantum Fields which are Manifestly Lorentz Invariant
have the inherent symmetries of the Complex Lorentz group and are TCP invariant by definition,
but for many Non-Relativistic and Effective Field Theories this automatic connection 
to TCP symmetry in the Complex Lorentz group is not there, and there needs to
be another process to derive and verify TCP symmetry.
The derivation of TCP using Variation of Action methods described here is an 
effort to provide that wider scope to test and verify TCP symmetry.  

The focus then of the key derivation in eqn (9) above is to show that
TCP invariance of the Lagrangian and endpoint variations
leads directly to TCP invariance of the Variation
of the Action and the overall Dynamical principle.
The other important factor is TCP invariance of the Measure.
In Schwinger's original derivations
of the Dynamical Principle and Variation of Actions methods,
the invariance of the measure is implicit, and is less significant an issue
here then TCP invariance of the Lagrangian and endpoint variations.

The invariance of the measure is a significant issue in Functional Path Integral
methods, and this leads to the most significant issue that comes from
proving TCP invariance of the variation of the Action in eqn(9) above.

The Action integral is the critical core also in
all Functional Path Integral derivation methods,
and there is a very close connection between the
Variation of Action method and Path Integral methods.
Dewitt shows (Chap.\ $10$ in \cite{Dewitt1}) that both
of these methods are derived from the same fundamental Action
and are corollary methods to each other. 

All the key elements shown in the proof here for
TCP invariance in Variation of the Action apply
as well to the Action in all Functional Path Integral methods.
In particular the strong condition tightly coupling T and C.
This TC condition applies to the Action at the core of all
Path Integral derivation methods as well,
and in principle TCP invariance as a wider Action principle
can be applied to Path Integral methods as well.
There are significant issues though in how Path Integral solutions
methods have developed which make it a non-trivial task to apply
the principle of TCP invariance in the Action derived here
to Path Integral solution methods.
This will be developed in more detail in future paper to follow.

\section{Simple Examples to Verify Proof}
\subsection{Simple Scalar, Vector, and Dirac Free Fields}
For the single Scalar free field case shown in Jost's Axiomatic proof,
and the three free field cases of Scalar, Vector, and Dirac free field
shown in the Pauli-Luder's proof, these three cases are
Manifestly Lorentz Invariant Relativistic Quantum Fields which
have TCP symmetry by definition from the Complex Lorentz group.
Then for these free Scalar, free Vector, and free Dirac quantum fields,
the variation of the Action will be invariant under
TCP by definition, and nothing in the Dynamical Principle or
Variation of Action methods changes this inherent TCP symmetry.
The results then for these simple free field cases comes
out identical to the original results in the Pauli-Luder's and Jost
proofs.

\subsection{Simple Bound State Case, Neutron $\beta$ Decay}
The $\beta$ decay of neutrons is one of the simplest and oldest types
of Bound State decay processes.
Inside a stable nucleus, the neutron can be a stable bound state of 3 quarks
with an indefinitely long lifetime. 
In an unstable nuclide, or as a free particle, the neutron becomes unstable and
undergoes $\beta^{-}$ decay with a mean lifetime under 882 secs. 

In a classic non-quark model, the neutron appears as a 
free particle that spontaneously performs the $\beta^{-}$ decay with no external interaction,
as shown in Fig 1. 
\begin{figure}[p]
\includegraphics[width=3.8in,scale=1.0]{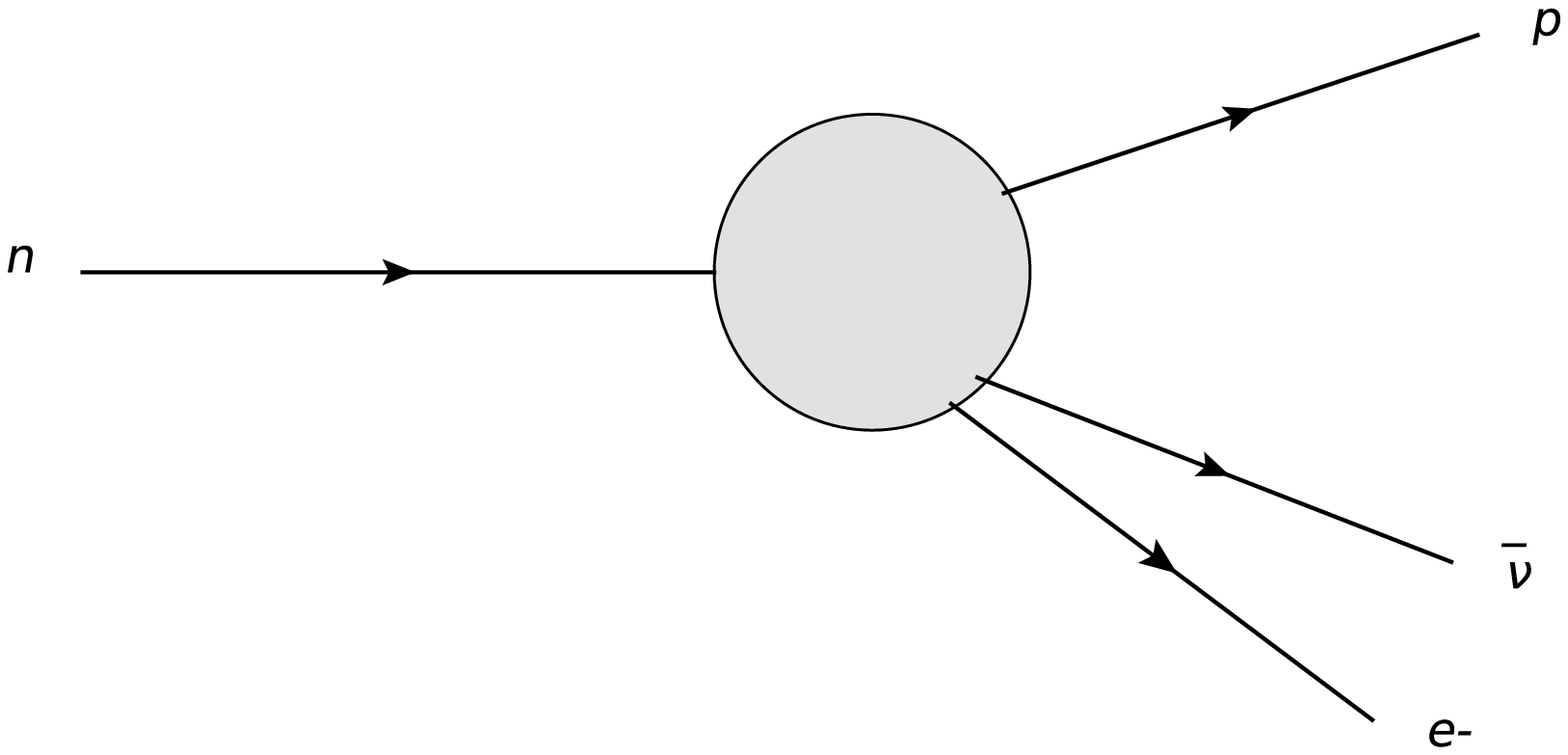}
\caption{Neutron $\beta$ decay as single free unstable particle.}
\end{figure}
In the S-matrix derivation method, this would be represented as an Unstable state
associated with Complex poles.

In reality, the current description of the neutron is a bound (or unstable) state of 3 quark constituents
$(u d d)$ in constant internal interaction with a swarm of gluons.
In this model, with constant internal interactions between the $(u d d)$ quark constituents
and gluons, the S-matrix method of derivation completely fails to handle 
the internal dynamics.

In its simplest form, using a lowest order spectator or free quark approximation model,
the 3 quark constituents are free particle fields which have limited interaction.
In $\beta$ decay,
only 1 d quark has a W weak decay vertex
while the other 2 quarks are spectators free of interaction.
The focus of the neutron $\beta^{-}$ decay then is 
the single $W^{-}$ weak decay vertex,
as shown in Fig 2.
\begin{figure}[p]
\includegraphics[width=3.8in,scale=1.0]{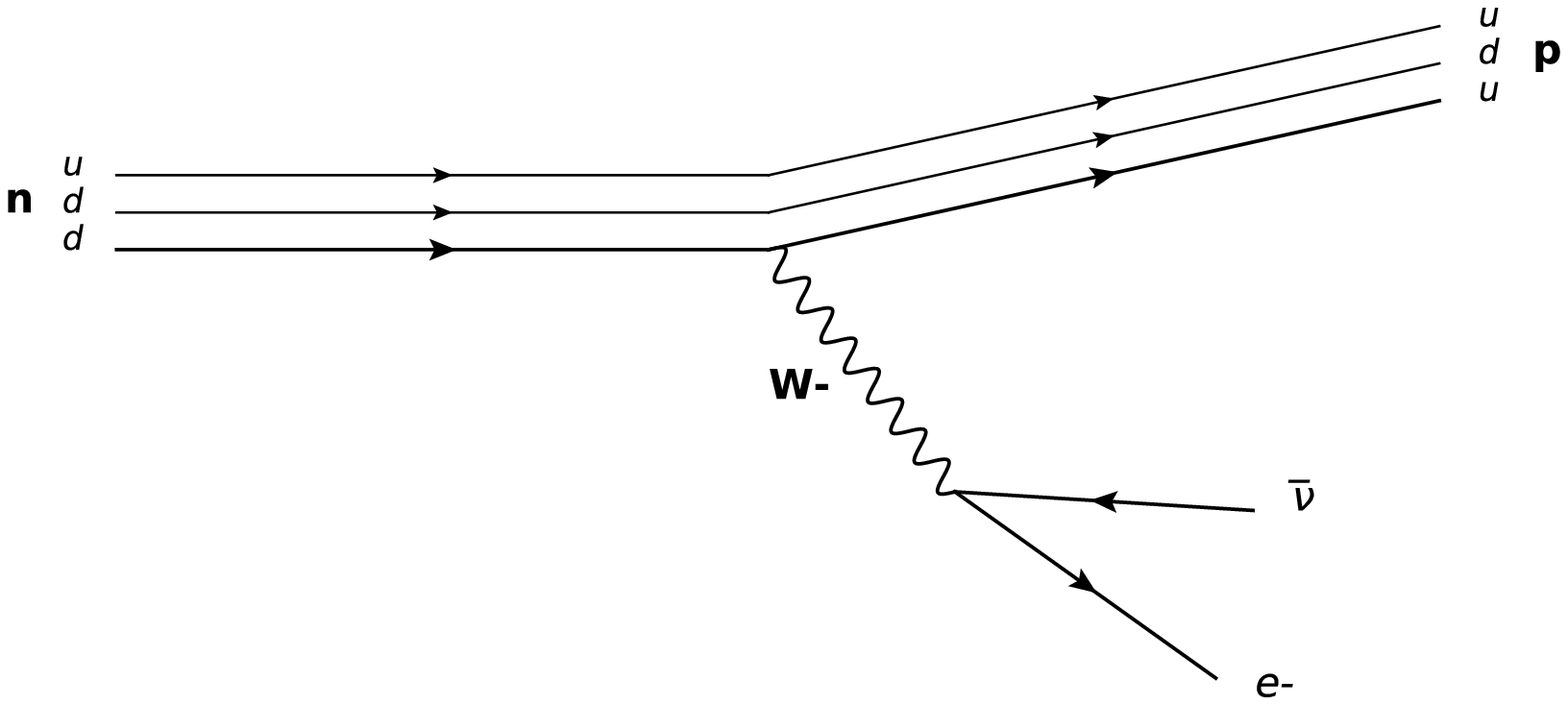}
\caption{Neutron $\beta$ Decay in free quark model}
\end{figure}

The Lagrangian for this system in its most abstract form consists of 
QCD interaction terms and Weak decay interaction terms:
\begin{displaymath}
   {\cal L} = {\cal L}_{qcd} + {\cal L}_{weak}
\end{displaymath}
The continous interaction of gluons with the 3 quark
constituents is in the  ${\cal L}_{qcd}$ term.
While the Weak decay interaction terms (below EW Symmetry breaking)
are in the ${\cal L}_{weak}$ term.

Since the gluon interactions are continuous short range internal interactions
that do not contribute to the $W$ decay vertex or the external decay products,
we can lower the importance of
these terms and focus primarily on the $W$ decay vertex on just one d quark.
This allows to use a free quark approximation to model the $\beta$ decay,
and focus on only one d quark. 

For the Weak decay interaction terms in ${\cal L}_{weak}$,
not all terms are needed here, only the charged current part for
the $W$ vertex:
\begin{equation}
  {\cal L}_{weak}
    = {\cal L}_{W} = - \frac{g}{2} \left[ \bar{u_{i}} \gamma^{\mu} \frac{1 - \gamma^{5}}{2} M_{ij}^{CKM} d_{j} 
                                      + \bar{\nu_{i}} \gamma^{\mu} \frac{1 - \gamma^{5}}{2} e_{i}
                              \right] W^{+}_{\mu} + h.c.
\end{equation}
This ${\cal L}_{W}$ set of terms is not invariant under P or TC, but is invariant under TCP.

Since the importance of the gluon terms in ${\cal L}_{qcd}$ has been minimized,
then the only terms in ${\cal L}$ that make significant contribution are the ${\cal L}_{W}$ terms
and:
\begin{displaymath}
  {\cal L} = {\cal L}_{qcd} + {\cal L}_{weak} = {\cal L}_{W}
\end{displaymath}
The test for TCP invariance here then depends only on ${\cal L}_{W}$ part of the
Lagrangian. Since the d quark here is a relativistic field, and ${\cal L}_{W}$ represents
Lorentz invariant terms, then  ${\cal L}$ here is TCP invariant.
Then the Action and Variation of the Action for the $\beta$ decay here is
invariant under TCP.

This example of $\beta$ decay of a 3 quark bound state was made
into simple 1 particle relativistic quark weak decay problem (in which case
TCP invariance is assumed by definition) by ignoring the interactions
of the gluons in ${\cal L}_{qcd}$. If the gluon interactions are
turned On though, this becomes a much more complicated 3-body unstable state
problem with both QCD and Weak decay terms. This cannot be handled
with simple Perturbation techniques on the relativistic Lagrangian,
but must be solved by other techniques, such as NRQCD effective field
theory techniques.
\section{Final Conclusions}
This is a response to the
challenge put out by Kobayashi and Sanda in 1992 \cite{Kobayashi1} that
the TCP/CPT proof needs to be updated beyond
the limits of the original Pauli-Luders and Jost proofs
(in particular the Asymptotic limit condition).
This paper represents
the first serious attempt to update TCP to handle 
more complicated Quantum fields including 
n-particle Bound states and Unstable states, and is
not limited to only Relativistic Quantum Fields where 
Manifest Lorentz Invariance and locality give TCP symmetry from
the Complex Lorentz group by definition.

The TCP proof using Action principles described in this paper are intended to expand
the scope of testing for TCP symmetry to a wider domain of Quantum Field cases,
including Non-Relativistic and Effective Field theories, where the connection
to inherent TCP symmetry in the Complex Lorentz group is not automatic.
This becomes a critical issue in many of the Effective Field theory techniques
used to solve Bound state and Unstable state problems such as NRQED,
NRQCD, and HQET, where the relativisitc Lagrangian (which has inherent TCP symmetry)
is transformed to a Non-relativsitc Hamiltonian or Lagrangian, and it is not
clear if the TCP symmetry of the original relativistic Lagrangian has been 
broken in the transformation process. 

This fundamental generalization of TCP to a wider domain of Quantum Field cases
was implicit in Schwinger's proof of Spin-Statistics in \cite{Schwinger-QFT2}
but not explicitly stated or proved.
At the time of this paper 
the development of the theory of QED bound states with the Bethe-Salpeter equation
was only in the very early stages, and the development of the theory of quark bound states
and QCD was over 15 years in the future.

On the key issue of the critical connection between T time inversion and C charge conjugation.
The relativistic wave equations for scalar (pseudo-scalar), vector (pseudo-vector), Dirac (Majorana),
and Maxwell wave equations, have both positive and negative energy solutions, and the
quantum field theories based on them have both particle and anti-particle solutions.
The use of finite dimensional non-unitary representations of the Lorentz group
have the property of passing from the positive energy particle solutions to their
anti-particle solutions by complex Lorentz transformations in a connected
set of transformations. But this property does not obtain for all unitary
representations of the Poincare group. This double covariance is the reason behind
all quantum field theories are in terms of the finite dimensional
representations of the Lorentz group being used in practice
\cite{Sudarshan1}.

The TCP proof described here can been directly applied to the
cases of 2-body and 3-body Bound States in QED and QCD,
but can also be extended to more complicated n-body Bound States
including both Relativistic and Non-Relatisitic Quantum field cases.
Many important applications of this method will be in
testing and verifying TCP in solutions of Bound state and Unstable
state problems using the NRQED, NRQCD, and HQET effective field theory
methods.

These more general and extended problems will
be developed in more detail in following papers.
In addition, the Action methods shown in this proof can be
extended to re-derive TCP using Functional Path Integral methods,
which will also be developed in following papers.

\end{document}